\documentclass[twocolumn]{aastex63}

\usepackage{amsmath}
\usepackage{gensymb}
\graphicspath{{./}{figures/}}
\usepackage{graphicx}
\usepackage{multirow}

\begin{document}

\title{Thermonuclear Explosions and Accretion-Induced Collapses of White Dwarfs in Active Galactic Nucleus Accretion disks}

\author[0000-0002-9195-4904]{Jin-Ping Zhu}
\affil{Department of Astronomy, School of Physics, Peking University, Beijing 100871, China; \url{zhujp@pku.edu.cn}}

\author[0000-0001-6374-8313]{Yuan-Pei Yang}
\affiliation{South-Western Institute for Astronomy Research, Yunnan University, Kunming, Yunnan, People’s Republic of China}

\author[0000-0002-9725-2524]{Bing Zhang}
\affiliation{Department of Physics and Astronomy, University of Nevada, Las Vegas, NV 89154, USA; \url{zhang@physics.unlv.edu}}

\author[0000-0002-8708-0597]{Liang-Duan Liu}
\affiliation{Department of Astronomy, Beijing Normal University, Beijing 100875, China}

\author[0000-0002-1067-1911]{Yun-Wei Yu}
\affiliation{Institute of Astrophysics, Central China Normal University, Wuhan 430079, China}

\author[0000-0002-3100-6558]{He Gao}
\affiliation{Department of Astronomy, Beijing Normal University, Beijing 100875, China}

\begin{abstract}

White dwarfs (WDs) embedded in gaseous disks of active galactic nucleus (AGNs) can rapidly accrete materials from the disks and grow in mass to reach or even exceed the Chandrasekhar limit. Binary WD (BWD) mergers are also believed to occur in AGN accretion disks. We study observational signatures from these events. We suggest that mass-accreting WDs and BWD mergers in AGN disks can lead to thermonuclear explosions that drive an ejecta shock breakout from the disk surface and power a slow-rising, relatively dim Type Ia supernova (SN). Such SNe Ia may be always outshone by the emission of the AGN disk around the supermassive black hole (BH) with a mass of $M_{\rm SMBH}\gtrsim 10^8\,M_\odot$. Besides, accretion-induced collapses (AICs) of WDs in AGN disks may occur sometimes, which may form highly-magnetized millisecond neutron stars (NSs). The subsequent spin-down process of this nascent magnetar can deposit its rotational energy into the disk materials, resulting in a magnetar-driven shock breakout and a luminous magnetar-powered transient. We show that such an AIC event could power a rapidly evolving and luminous transient for a magnetic field of $B\sim10^{15}\,{\rm G}$. The rising time and peak luminosity of the transient, powered by a magnetar with $B\sim10^{14}\,{\rm G}$, are predicted to have similar properties with those of superluminous supernovae. AIC events taking place in the inner parts of the disk around a relatively less massive supermassive BHs ($M_{\rm SMBH}\lesssim10^8\,M_\odot$) are more likely to power the transients that are much brighter than the AGN disk emission and hence easily to be identified.

\end{abstract}


\section{Introduction} \label{sec:intro}

Massive stars are believed to exist in the accretion disks of active galactic nuclei (AGNs). Such AGN stars can be either the result of in situ formation inside the accretion disk or be captured from the nuclear star clusters around the AGNs \citep[e.g.,][]{artymowicz1993,collin1999,goodman2003,goodman2004,wang2011,wang2012,fabj2020,cantiello2021}. These AGN stars will end up with supernovae (SNe) which can eject heavy elements into the disk and hence, offer a possible explanation for the observational features of {high metallicity} environments in AGN disks \citep[e.g.,][]{artymowicz1993,hamann1999,warner2003}. Some compact objects, including white dwarfs (WDs), neutron stars (NSs), and black holes (BHs), can be thus formed within AGN disks. These compact objects can also be captured from the surrounding nuclear star clusters. The disk of an AGN provides a natural environment for stars and compact objects to accrete materials and to migrate within it \citep[e.g.,][]{mckernan2012,yang2020,dittmann2021,jermyn2021,wang2021,tagawa2021,kimura2021}. Some of these stars can be very massive and have high spin caused by accretion \citep{dittmann2021,jermyn2021}, so that they can easily produce high-spin stellar remnants. {Abundant compact objects, especially with the presence of the massive BHs ($M>10M_\odot$), would likely accrete, collide, and merge within the trapping orbits, and hence, would grow into $\sim100\,M_\odot$ intermediate mass BHs \citep{mckernan2012,secunda2019,yang2019}.} Some AGN stars can be tidally disrupted by these BHs that can power micro-tidal disruption events \citep{yang2021}. The death of high-spin stars and neutron star mergers are expected to power gamma-ray burst (GRB) jets, which would be always choked by the dense atmosphere of the disks \citep{zhu2021b,zhu2021a,perna2021a}. \cite{zhu2021b} suggested that these choked jets can produce high-energy neutrinos that may contribute a substantial fraction of the diffuse neutrino background. A candidate electromagnetic (EM) counterpart emerged from an AGN, explained as ram-pressure stripping of gas within the kicked BH hill sphere colliding with the AGN disk gas \citep{mckernan2019}, was reported by the Zwicky Transient Facility \citep{graham2020} which was thought to be associated with a $(85 + 66)\,M_\odot$ binary BH merger (GW190521) detected by the LIGO/Virgo collaboration \citep{abbott2020}. This connection provided plausible evidence of a potentially important AGN channel for compact star mergers. 

The classical understanding of the Type Ia SN progenitor is that they are powered by thermonuclear explosions of mass-accreting WD, either in a binary with a non-degenerate (i.e., a main-sequence star, a red-giant star, or a helium star) star companion or in a binary WD (BWD) system \citep[e.g.][]{whelan1973,nomoto1982,iben1984,webbink1984,nomoto1984,hachisu1996,hillebrandt2000,leigh2020}. In an AGN disk environment, it is interesting to note that WDs embedded in AGN disks can in principle grow and spin up via accretion from the disk materials, which may power thermonuclear explosions. BWD mergers are also expected to occur in AGN disks \citep[e.g.,][]{cheng1999,mckernan2020}. The EM signatures of these AGN-WD-powered thermonuclear explosions could show some unique properties compared with classical Type Ia SNe. 

Moreover, mass-accreting WDs and BWD mergers\footnote{During the merger of a BWD, both mass and angular momentum would be retained in the massive WD merger product.} may result in the formation of NSs through accretion-induced collapse \citep[AIC; e.g.,][]{canal1976,miyaji1980,sajo1985,nomoto1991,schwab2016}. The NSs formed by the AIC of WDs could rotate rapidly with spin periods of only a few milliseconds \citep{dessart2006} and potentially have strong magnetic fields which could be significantly amplified via magnetorotational instability \citep{duncan1992,dessart2007,cheng2014}. Some bright transients, e.g., superluminous supernovae \citep[SLSNe; e.g.,][]{kasen2010,woosley2010,metzger2014a,wang2015,yu2017,liu2017,nicholl2017}, mergernovae \citep[magnetar-powered kilonovae;][]{yu2013,metzger2014b}, X-ray/optical afterglow powered by dissipation of magnetar winds from binary NS (BNS) mergers \citep{zhang2013,sun2017,sun2019,xue2019}, and optical transients from classical AIC of WDs \citep{yu2015,yu2019a,yu2019b}, have long been proposed to involve spin-down of a rapidly spinning magnetar which continuously injects energy into the system. Therefore, it is expected that subsequent spin-down of the nascent millisecond magnetar formed from an AIC within a AGN disk could potentially power luminous explosions as well. 

In this letter, we perform an investigation of the EM properties of thermonuclear explosions and AICs of WDs in the dense environment of an AGN disk\footnote{As this work was being finalized, we noticed the work by \cite{perna2021b}. Different from our work that focuses on the EM signatures of AIC of WDs in AGNs, they discussed the event rates and observable prospects of AIC of NSs to BHs in AGNs.}.

\section{Models} \label{sec:model}

We discuss the physics of two scenarios: (1) thermonuclear explosion of an accreting WD in an AGN disk and (2) a WD AIC and the subsequent transient powered by a millisecond magnetar produced by the AIC event. WDs within the AGN disks are expected to be located in the trapping orbit \citep[e.g.,][]{bellovary2016,tagawa2020}, i.e., $a \simeq 10^3 R_{\rm SMBH}$ from the SMBH, where $R_{\rm SMBH} \equiv GM_{\rm SMBH}/c^2$ is the radius of SMBH, $G$ is the gravitational constant, $M_{\rm SMBH}$ is the mass of SMBH, and $c$ is the speed of light. For a typical AGN with $M_{\rm SMBH} = 10^8\,M_\odot$ based on AGN observations \citep[e.g.,][]{woo2002,kollmeier2006}, the density $\rho_0$ in the disk mid-plane  and the disk scale height $H$ are $\rho_0 \sim 1\times 10^{-9}\,{\rm cm}$ and $H\sim2\times10^{14}\,{\rm cm}$, respectively, for the disk model by \cite{sirko2003}; and they are $\rho_0 \sim 4 \times 10^{-13}\,{\rm g}\,{\rm cm}^{-3}$ and $H \sim 6 \times 10^{14}\,{\rm cm}$, respectively, for the disk model by \cite{thompson2005}. We simply set median values of $\rho= 4\times10^{-11}\,{\rm g}\,{\rm cm}^{-3}$ and $H \sim 4\times 10^{14}\,{\rm cm}$ of these two specific disk models to assess the observational features of thermonuclear explosions and AIC events within an AGN disk. The vertical density profile can be described by an isothermal atmosphere model \citep[e.g.,][]{netzer2013}, i.e., $\rho(a,z) = \rho_0(a)\exp[-z^2/2H^2(a)]$. For simplicity, we approximately adopt a uniform density profile $\rho \approx \rho_0$ for $z < H$, since $\rho$ decrease rapidly at $z>H$ for a disk with an exponentially decaying density profile.

\subsection{Scenario I: Thermonuclear Explosion} \label{sec:sce:I}

In view that the AGN stars could have near-critical rotation rates \citep{jermyn2021}, it is reasonable to assume that the spins of the remnant WDs in the AGN disks are extremely high. Such a high-spin WD could have a larger mass than a low-spin one through accretion, which could be even up to $\gtrsim 2\,M_\odot$, finally resulting in a super-Chandrasekhar SNe Ia  \citep[e.g.,][]{hachisu2012,wang2014,benvenuto2015}. An SN Ia with a total mass greater than the Chandrasekhar mass limit is usually thought to result from merger of a BWD \citep[see][for a review]{maoz2014}. Observations of super-Chandrasekhar SNe Ia showed a large amount of radioactive nickel \citep[e.g.,][]{mccully2014,scalzo2014}. Some extreme events can even produce $\sim 1.8\,M_\odot$ of $^{56}{\rm Ni}$ \citep[e.g.,][]{silverman2011,taubenberger2011}. Hereafter, for those SNe Ia occurring in AGN disks, we conservatively set the mass of the ejecta as $M_{\rm ej} = 1.5\,M_\odot$, with $M_{\rm Ni} = 1\,M_\odot$ for rough estimations. The initial velocity of the SNe Ia ejecta is suggested to be of the order of $10^4\,{\rm km}\,{\rm s}^{-1}$ based on the observational optical spectral features in many SNe Ia  \citep[e.g.,][]{parrent2012,pereira2013,maoz2014}, so we assume the initial velocity of the ejecta is $v_{\rm i} = 20,000\,{\rm km}\,{\rm s}^{-1}$. 

After an thermonuclear explosion, the ejecta would expand, sweep the AGN disk material, form a radiation-mediated shock, and finally break out from the accretion disk of the AGN. The shock breakout can take place when the photons ahead of the shock diffuse faster than shock propagation, i.e., the diffusion timescale $t_{\rm bo,diff}$ is close to the shock propagation time $t_{\rm bo,exp}$, where $t_{\rm bo,diff}\approx\kappa\rho_0(H-h_{\rm bo})^2/c$ and $t_{\rm bo,exp} \approx (H - h_{\rm bo})/v_{\rm bo}$ with $h_{\rm bo}$ being the vertical location when photons break out, and $v_{\rm bo}$ is the breakout velocity of the shock. $\kappa = 0.34 \ {\rm cm}^{2} \ {\rm g}^{-1}$ is adopted as the opacity for both the ejecta and the disk atmosphere. {Hereafter, the convention $Q_x = Q/10^x$ is adopted in cgs units. } Therefore, the distance between the location of shock breakout and disk surface can be calculated as
\begin{equation}
\label{eq:distance}
    d_{\rm bo} = H - h_{\rm bo} \approx 2.4\times 10^{12}\,{\rm cm}\,\rho_{0,-10.4}^{-1}v_{\rm bo,9}^{-1}.
\end{equation}
One can conclude that $d_{\rm bo}\ll H$, which means that the location of shock breakout is close to the AGN disk surface.

When the shock breaks out from the disk, the total swept mass by the ejecta can be estimated as
\begin{equation}
    M_{\rm sw} \approx \frac{4\pi \rho_0 H^3}{3} = 5.4\,M_\odot\,\rho_{0,-10.4}H_{14.6}^3,
\end{equation}
and the shock breakout velocity of the ejecta is
\begin{equation}
\begin{split}
    v_{\rm bo} &\approx \sqrt{M_{\rm ej}v_{\rm i}^2/(M_{\rm ej} + M_{\rm sw})} \\
    &\approx 0.03\,c \,M_{{\rm ej},0.2\odot}^{1/2}M_{{\rm tot},0.8\odot}^{-1/2}v_{{\rm i},9.3},
\end{split}
\end{equation}
where $M_{\rm tot} = M_{\rm ej} + M_{\rm sw}$. By comparing the radiation energy density and the kinetic energy density with an adiabatic index  $\gamma = 4 / 3$, the downstream blackbody temperature of the ejecta shock when it breaks out can be expressed as $T_{\rm bo}^{\rm BB} = (7\rho_0v_{\rm bo}^2/2a)^{1/4} \approx 3.6\times10^5\,{\rm K}\,\rho_{0,-10.4}^{1/4}M_{{\rm ej},0.2\odot}^{1/4}M_{{\rm tot},0.8\odot}^{-1/4}v_{{\rm i},9.3}^{1/2}$, where $a$ is the radiation constant. Note that $T_{\rm bo}^{\rm BB}$ is the temperature at $\tau \approx c / v_{\rm bo}$, not the photosphere temperature at $\tau \approx 1$ that observers would see. The latter is given by
\begin{equation}
\begin{split}
    T_{\rm bo}^{\rm obs} &\approx \left( \frac{v_{\rm bo}}{c} \right) ^ {1/4}T_{\rm bo}^{\rm BB} \\ 
    &\approx 1.5\times10^5\,{\rm K}\,\rho_{0,-10.4}^{1/4}M_{{\rm ej},0.2\odot}^{3/8}M_{{\rm tot},0.8\odot}^{-3/8}v_{{\rm i},9.3}^{3/4}.
\end{split}
\end{equation}
{The total energy of the breakout photons is approximately $E_{\rm bo} \approx \rho_0\pi H^2 d_{\rm bo}v_{\rm bo}^2 \approx 4.1\times 10^{49}\,{\rm erg}\,H_{14.6}^2M_{{\rm ej},0.2\odot}^{1/2}M_{{\rm tot},0.8\odot}^{-1/2}v_{{\rm i},9.3}$, where $d_{\rm bo} \approx 4.0\times10^{12}\,{\rm cm}\,\rho_{0,-10.4}^{-1}M_{{\rm ej},0.2\odot}^{-1/2}M_{{\rm tot},0.8\odot}^{1/2}v_{\rm bo,9.3}^{-1}$ is obtained based on Equation (\ref{eq:distance}).} This energy is released in a timescale of
\begin{equation}
\begin{split}
    t_{\rm bo,diff} &\approx \kappa\rho_0d_{\rm bo}^2 / c \\ 
    &\approx 0.7\,{\rm hr}\,\rho_{0,-10.4}^{-1}M_{{\rm ej},0.2\odot}^{-1}M_{{\rm tot},0.8\odot}v_{\rm bo,9.3}^{-2},
\end{split}
\end{equation}
so that the luminosity of the ejecta breakout can be estimated as
\begin{equation}
\begin{split}
    &L_{\rm bo} \approx \frac{E_{\rm bo}}{t_{\rm bo,diff}} \\
    & \approx 1.6\times 10^{46}\,{\rm erg}\,{\rm s}^{-1}\,\rho_{0,-10.4}H_{14.6}^2M_{{\rm ej},0.2\odot}^{3/2}M_{{\rm tot},0.8\odot}^{-3/2}v_{{\rm i},9.3}^3.
\end{split}
\end{equation}

The radiated emission from a subsequent SN would be powered by the energy released via the radioactive decay of $^{56}{\rm Ni}$ and $^{56}{\rm Co}$. The expression for the radioactivity luminosity is \citep{nadyozhin1994,arnett1996}
\begin{equation}
    L_{\rm r} = (6.45e^{-t/t_{\rm Ni}} + 1.45e^{-t/t_{\rm Co}})\frac{M_{\rm Ni}}{1M_{\odot}}\times10^{43}\,{\rm erg}\,{\rm s}^{-1},
\end{equation}
where $t_{\rm Ni} = 8.8\,{\rm day}$ and $t_{\rm Co} = 111.3\,{\rm day}$ are the mean lifetimes of the nuclei, and $M_{\rm Ni}$ is in solar masses. The brightness of the SN increases until the expanded ejecta and the swept material are as old as the effective diffusion timescale, i.e., 
\begin{equation}
\begin{split}
    t_{\rm sn,diff} & \approx \left[ \frac{3\kappa(M_{\rm ej} + M_{\rm sw})}{4\pi v_{\rm ej}c} \right] ^ {1/2} \\
    &\approx 73\,{\rm day}\,M_{{\rm ej},0.2\odot}^{-1/4}M_{{\rm tot},0.8\odot}^{3/4}v_{{\rm i},9.3}^{-1/2},
\end{split}
\end{equation}
where we roughly take $v_{\rm bo}$ as the final characteristic ejecta velocity $v_{\rm ej}$. Based on the general expression of ``Arnett's law" \citep{arnett1979}, the peak radiated luminosity would equal the instantaneous radioactivity luminosity at the peak time $t_{\rm sn,peak}\approx t_{\rm sn,diff}$. The peak luminosity of the SN is then
\begin{equation}
    L_{\rm sn,peak} \approx L_{\rm r}(t_{\rm sn,peak}) \approx 7.5\times 10^{42}\,{\rm erg}\,{\rm s}^{-1},
\end{equation}
with an effective temperature 
\begin{equation}
    T_{\rm sn,eff} = \left( \frac{L_{\rm sn,peak}}{4\pi\sigma_{\rm SB} v_{\rm ej}^2t_{\rm sn, diff}^2} \right) ^ {1/4} \approx 4.2\times 10^3\,{\rm K},
\end{equation}
where each parameter is its typical value given above and $\sigma_{\rm SB}$ is the Stefan-Boltzmann constant.

\subsection{Scenario II: AIC of WDs} \label{sec:sce:II}

A millisecond, highly magnetized NS may be born directly via AIC of a massive WD. The thermonuclear explosion may be avoided in this case. This is another scenario for the aftermath of accreting WDs embedded in AGN disks.

In the standard picture of SNe, the total gravitational binding energy released in the collapse of the core of an NS is approximately  $E_{\rm gr} \approx 3GM_{\rm ns}^2/5R_{\rm ns} \approx 3.6\times10^{53}\,{\rm erg}\,(M_{\rm ns} / 1.5\,M_\odot)^2R_{\rm ns,6}^{-1}$.  Most of this energy is radiated in the forms of neutrinos \citep{colgate1966}. The kinetic energy of the SN ejecta only amount to $\lesssim1\%$ of the gravitational energy. Similarly, for AIC of an WD to form a proto-NS, huge gravitational energy is released and carried away by neutrinos. However, the neutrino mechanism to power an SN may not work effectively within a collapsing WD, and only a weak explosion with energy to be lower than $10^{50}\,{\rm erg}$ is expected \citep{dessart2006}. Neutrinos would escape freely due to the relatively low density of the AGN disk. 

The end prodict of a WD AIC is likely a magnetar. After its formation, a magnetar-driven wind can create a high-pressure bubble of the magnetic field and relativistic particles that can sweep and drive a shock into the AGN materials \citep{kasen2016,suzuki2017}. This shock can finally break out from the surface of the AGN disk, which is a process similar to the magnetar-driven shock breakouts discussed for SLSNe \citep{kasen2016,liu2021} and mergernovae \citep{li2016}. 

Given by some AIC simulations, the spin of the nascent NS of two specific models shown in \cite{dessart2006} are $2.2\,{\rm ms}$ and $6.3\,{\rm ms}$, respectively, while the surface magnetic field of the AIC NS could be as high as $\sim10^{15}\,{\rm G}$ \citep{dessart2007}. NSs formed via BWD mergers can have rotational periods mainly in the range of $\sim (1-7)\,{\rm ms}$ \citep{schwab2021}. Motivated by these simulation results, the initial period $P_{\rm i} = 2\,{\rm ms}$ and magnetic field $B = 10^{14}\,{\rm G}$ for the magnetar are adopted. Moreover, we set the NS mass to be $M_{\rm ns} = 1.5\,M_\odot$ and the NS radius to be $R_{\rm ns} = 10\,{\rm km}$. The total rotational energy released during the spin-down of the magnetar is $E_{\rm tot} \approx E_{\rm rot} \approx I_{\rm ns}\Omega_{\rm i}^2 / 2 \approx 5.9\times 10^{51}\,{\rm erg}\,P_{\rm i,-2.7}^{-2}$, where $I_{\rm ns} = 1.2\times10^{45}\,{\rm g}\,{\rm cm}^2$ and $\Omega_{\rm i} = 2\pi / P_{\rm i}$. If the magnetar spindown mechanism is dominated by  magnetic dipole radiation, the spin-down luminosity is given by
\begin{equation}
    L_{\rm sd}(t) = \frac{L_{\rm sd,i}}{(1 + t/t_{\rm sd})^2} \approx \left\{\begin{matrix}
    L_{\rm sd,i}\ &{\rm if}\ t<t_{\rm sd}, \\
    L_{\rm sd,i}/t_{\rm sd}^2\ &{\rm if}\ t>t_{\rm sd},
\end{matrix}\right.
\end{equation}
with the spindown timescale of $t_{\rm sd} = 6I_{\rm ns}c^3/(B^2R_{\rm ns}^6\Omega_{\rm i}^2) = 23\,{\rm day}\,B_{14}^{-2}P_{\rm i,-2.7}^2$, and the initial spindown luminosity of $L_{\rm sd,i} = E_{\rm rot} / t_{\rm sd} = 3.0\times10^{45}\,{\rm erg}\,{\rm s}^{-1}\,B_{14}^2P_{\rm i,-2.7}^4$.

One can estimate the breakout velocity of the magnetar-driven shock as $v_{\rm bo} \approx \sqrt{2L_{\rm sd,i}t_{\rm bo,pro}/M_{\rm sw}}$, where the time for the shock to propagate within the disk before breakout is $t_{\rm bo,pro} \approx H/\sqrt{2}v_{\rm bo}$. Thus, the shock breakout velocity is
\begin{equation}
    v_{\rm bo} / c \approx 0.018\, \rho_{0,-10.4}^{-1/3} H_{14.6}^{-2/3}B_{14}^{2/3}P_{\rm i,-2.7}^{4/3}.
\end{equation}
while $t_{\rm bo,pro} \approx 6.0\,{\rm day}\,\rho_{0,-10.4}^{1/3} H_{14.6}^{5/3}B_{14}^{-2/3}P_{\rm i,-2.7}^{-4/3}$. Similar to shock breakout by thermonuclear explosion, we can give the total energy of the magnetar-driven shock breakout photons
\begin{equation}
    E_{\rm bo} \approx 2.4\times10^{49}\,{\rm erg}\,\rho_{0,-10.4}^{-1/3} H_{14.6}^{4/3}B_{14}^{2/3}P_{\rm i,-2.7}^{4/3},
\end{equation}
the observed temperature
\begin{equation}
    T_{\rm bo}^{\rm obs} \approx 9.9\times 10^{4}\,{\rm K}\,H_{14.6}^{-1/2}B_{14}^{1/2}P_{\rm i,-2.7},
\end{equation}
the diffusion timescale
\begin{equation}
    t_{\rm bo,diff} \approx 2.1\,{\rm hr}\,\rho_{0,-10.4}^{-1/3} H_{14.6}^{4/3}B_{14}^{-4/3}P_{\rm i,-2.7}^{-8/3},
\end{equation}
and the luminosity
\begin{equation}
    L_{\rm bo} \approx \frac{E_{\rm bo}}{t_{\rm bo,diff}} \approx 3.2\times10^{45}\,{\rm erg}\,{\rm s}^{-1}\,B_{14}^{2}P_{\rm i,-2.7}^{4}.
\end{equation}

After the magnetar-driven shock breaks out from the surface of the AGN disk, the swept disk material would become the ``ejecta" of this SN-like explosion (i.e., $M_{\rm ej} \approx M_{\rm sw}$). The energy injection from the magnetar spindown would still continue. Most of the rotational energy would be transformed into the kinetic energy of the ejecta within the spindown timescale $t_{\rm sd}$, so that we can take the final characteristic value of the ejecta velocity as
\begin{equation}
    v_{\rm ej} \approx \sqrt{2E_{\rm rot}/M_{\rm ej}} \approx 0.035\,c\,\rho_{0,-10.4}^{-1/2}H_{14.6}^{-3/2}P_{\rm i,-2.7}^{-1}.
\end{equation}
The effective diffusion time is then
\begin{equation}
    t_{\rm sn,diff} = \left( \frac{3\kappa M_{\rm ej}}{4\pi v_{\rm ej}c} \right)^{1 / 2} \approx 61\,{\rm day}\,\rho_{0,-10.4}^{3/4}H_{14.6}^{9/4}P_{\rm i,-2.7}^{1/2},
\end{equation}
which is larger than the spindown timescale ($t_{\rm sn,diff} > t_{\rm sd}$). The peak luminosity can be estimated based on ``Arnett's law" similar to the calculations in the Section \ref{sec:sce:I}, i.e.,
\begin{equation}
\begin{split}
    L_{\rm sn,peak} &\approx L_{\rm sd}(t_{\rm sn,diff}) \approx L_{\rm sd,i}t_{\rm sd}^2/t_{\rm sn,diff}^2 \\
    &\approx 2.2\times 10^{44}\,{\rm erg}\,{\rm s}^{-1}\,\rho_{0,-10.4}^{-3/2}H_{14.6}^{-9/2}B_{14}^{-2}P_{\rm i,-2.7}^{7}.
\end{split}
\end{equation}
The effective temperature is calculated as
\begin{equation}
    T_{\rm sn,eff} = 1.0\times10^4\,{\rm K}\,\rho_{0,-10.4}^{-1/2}H_{14.6}^{-3/2}B_{14}^{-1/2}P_{\rm i,-2.7}^{2}.
\end{equation}
We note that energy ejection from the neutrino mechanism is ignored in the calculation, since its energy is much lower than the energy released by the magnetar before shock breakout. Furthermore, the collapse of WD and the disk outflow may lead to a $\sim0.001-0.1\,M_\odot$ of ejecta with velocity $\sim0.1\,c$ \citep[e.g.,][]{dessart2006,dessart2007,metzger2009}. The relatively small mass of the AIC ejecta has no significant effect on the results of the above calculation so we also do not take it into account.

In summary, both scenarios for the aftermath of a massive accreting WD within the trapping orbit of a typical AGN disk can produce a bright shock breakout emission signal in the UV band and an optical SN-like explosion. The shock breakout for a WD ending up with a thermonuclear explosion has a luminosity of the order of $10^{46}\,{\rm erg}\,{\rm s}^{-1}$ with a duration of $\sim 0.7\,{\rm hrs}$. Subsequent emission, with a peak luminosity of $\sim 8\times 10^{42}\,{\rm erg}\,{\rm s}^{-1}$ and a peak time of $\sim 73\,{\rm days}$, would be powered by the radioactive decay of $^{56}{\rm Ni}$ and $^{56}{\rm Co}$. For a classical SN Ia, the lightcurve would rise to maximum within $\sim15-20\,{\rm days}$ with a typical luminosity of $\sim1.7\times10^{43}\,{\rm erg}\,{\rm s}^{-1}$ \citep[e.g.,][]{maoz2014}. Thus, the Ia SN explosion in AGN disk could have a much slower rise and dimmer peak luminsoity compared with a classical SN Ia. For the other scenario, the magnetar-driven shock breakout can have a luminosity of $\sim3\times10^{45}\,{\rm erg}\,{\rm s}^{-1}$ and a duration of $\sim2\,{\rm hrs}$. The main peak is driven by the diffusion of radiation from magnetar heating, which highly depends on the initial spin and magnetic field of the nascent magnetar. For $P_{\rm i} = 2\,{\rm ms}$ and $B = 10^{14}\,{\rm G}$ with an assumption of magnetic dipole radiation dominating the spindown, the peak luminosity and peak time of the magnetar-powered transient are respectively $\sim2\times10^{44}\,{\rm erg}\,{\rm s}^{-1}$ and $\sim60\,{\rm days}$, which is comparable to the observable properties of an SLSN \citep{nicholl2015,nicholl2017,yu2017,lunnan2018}. Thus, an AIC event in AGN disks can potentially power an SLSN-like transient. 

\section{Detectability of the Transients}

\subsection{Parameters Dependencies and Comparison with the AGN Disk Spectra}

\begin{figure*}[htbp]
    \centering
	\includegraphics[width = 0.49\linewidth , trim = 55 0 60 20 , clip]{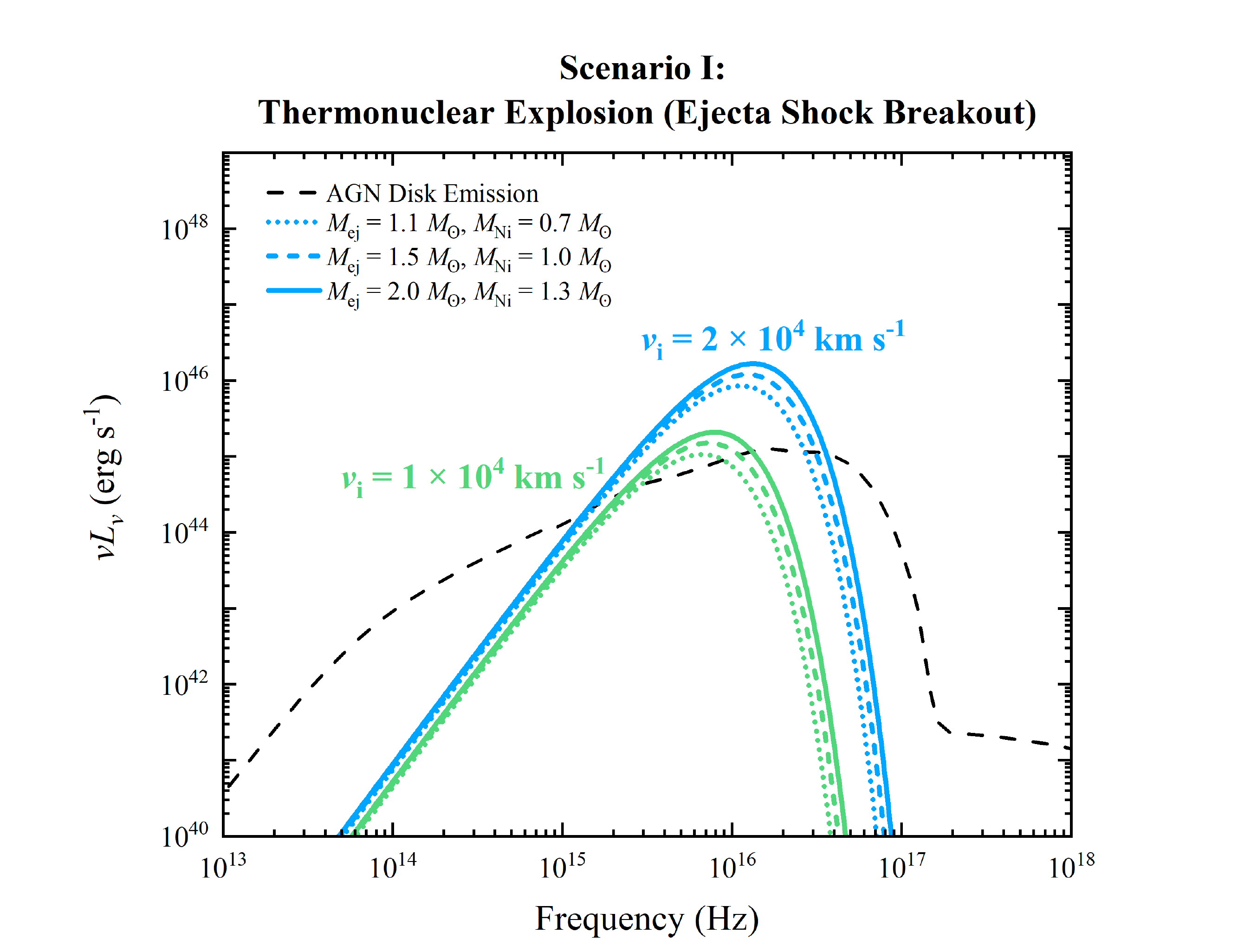}
	\includegraphics[width = 0.49\linewidth , trim = 55 0 60 20 , clip]{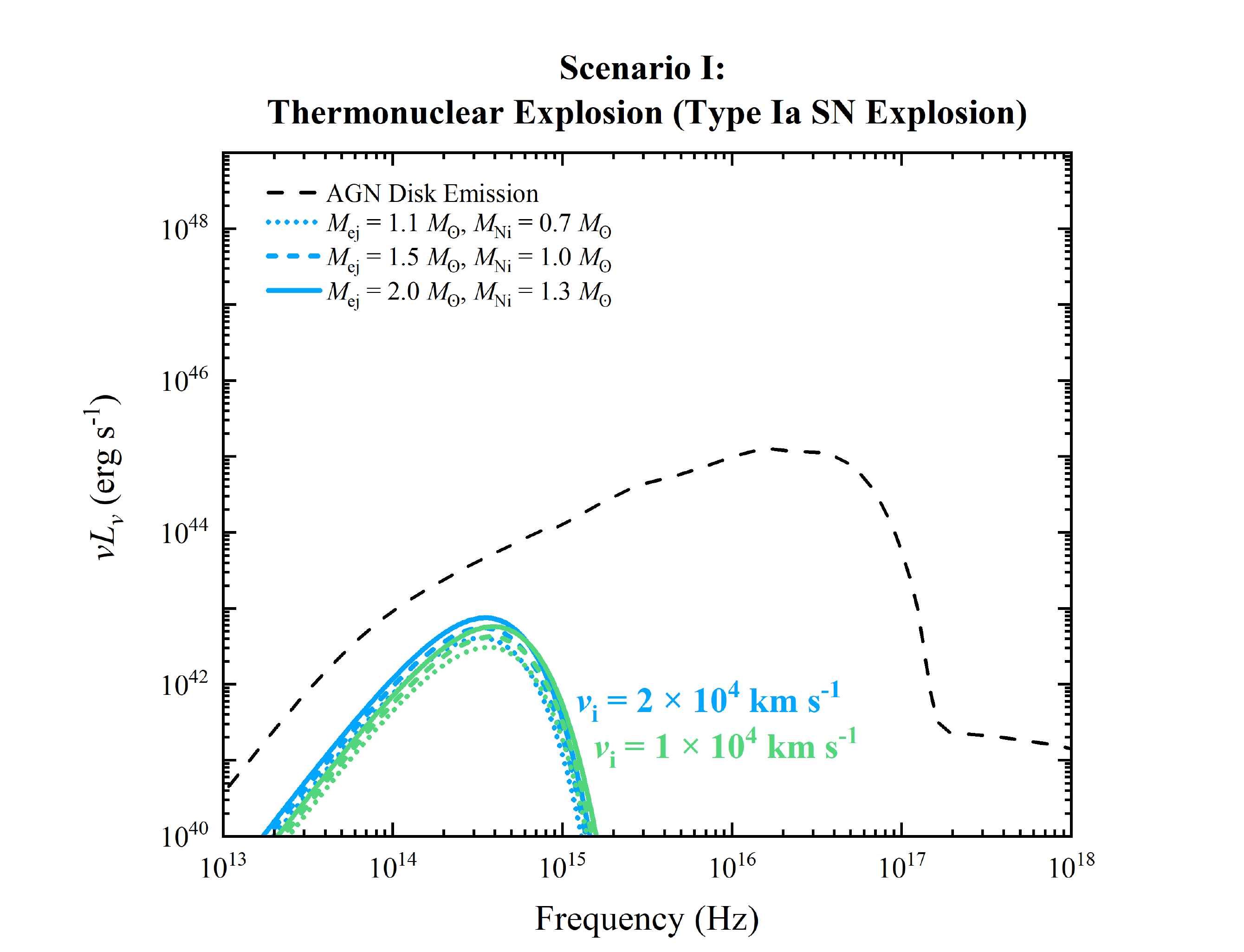}
	\includegraphics[width = 0.49\linewidth , trim = 55 0 60 20 , clip]{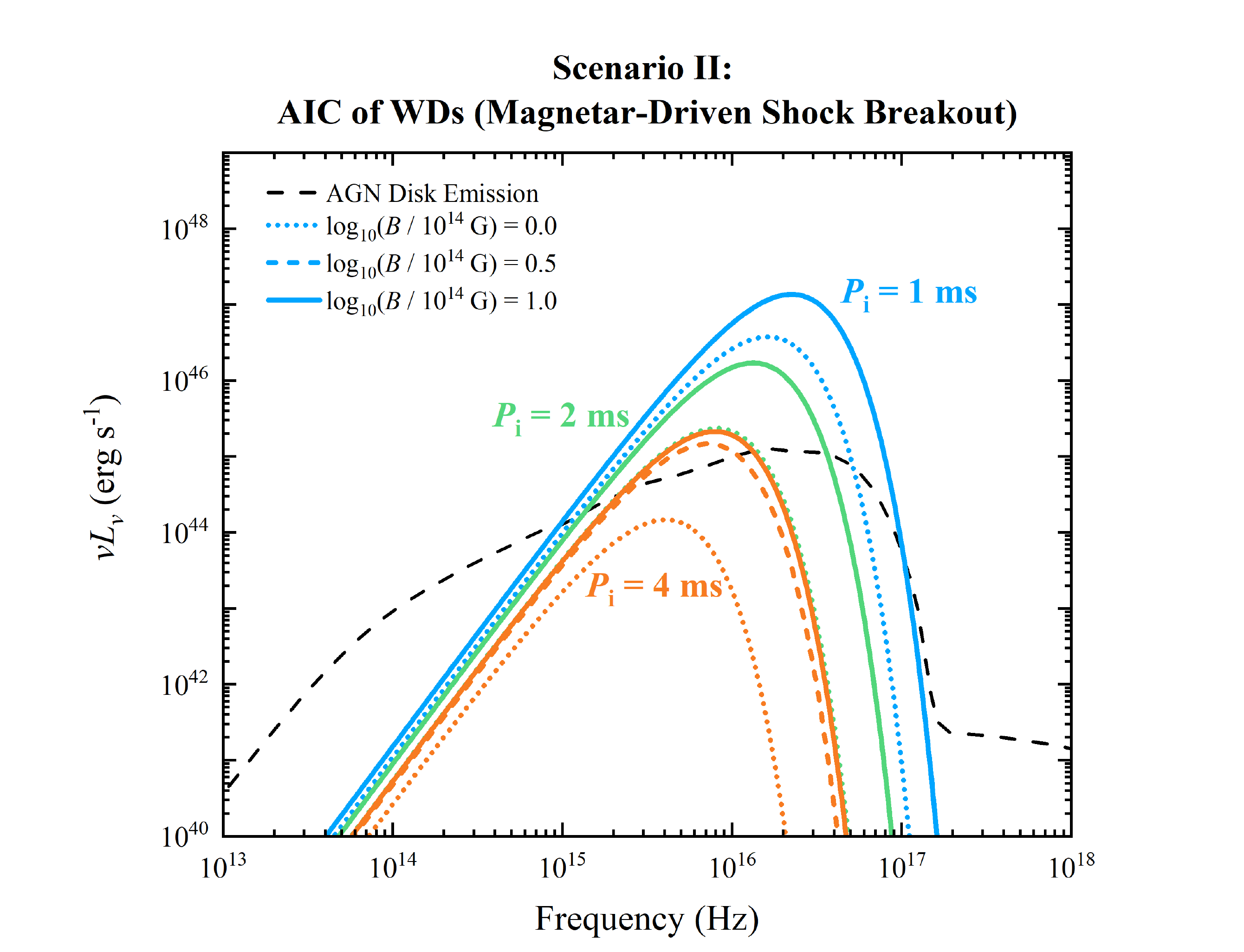}
	\includegraphics[width = 0.49\linewidth , trim = 55 0 60 20 , clip]{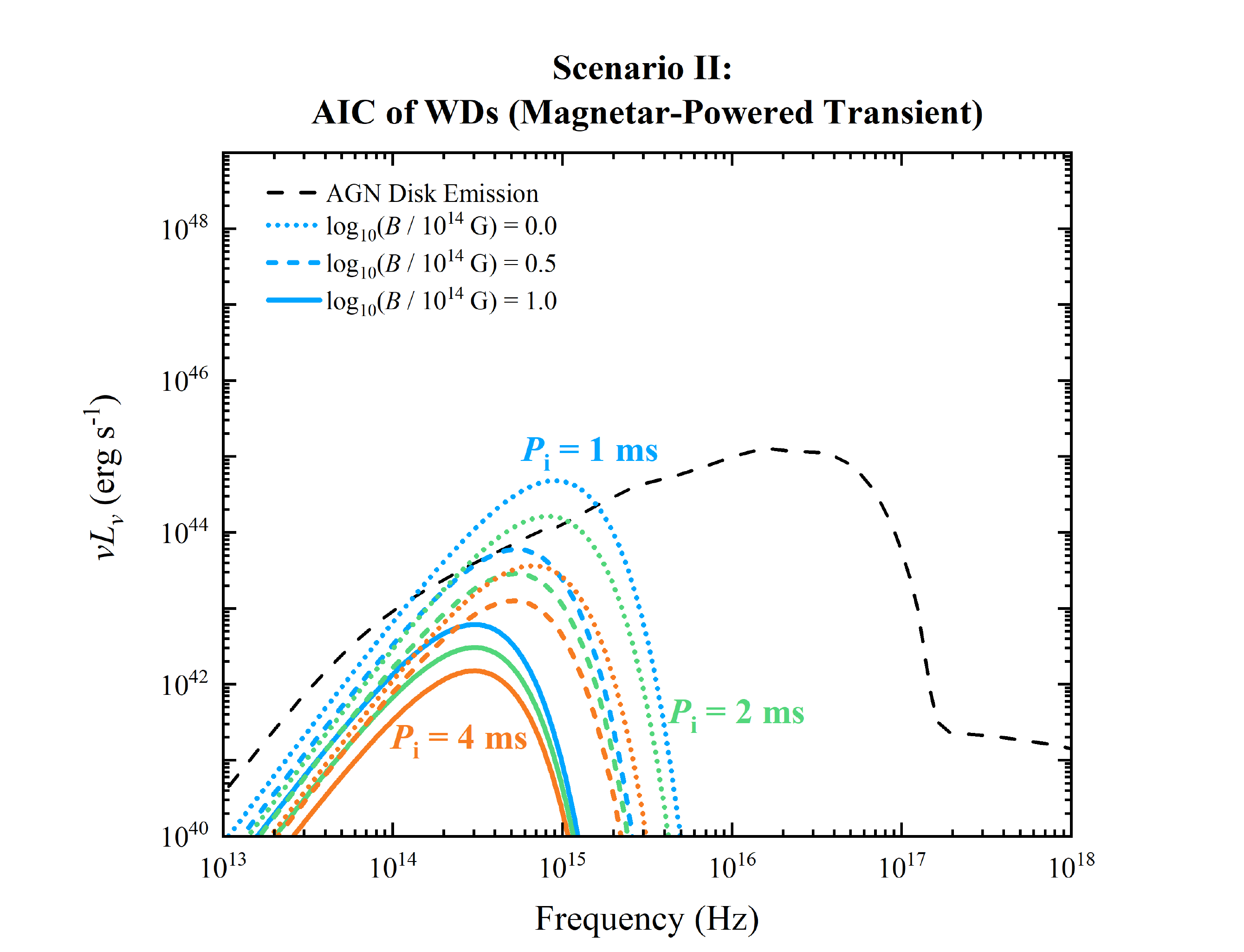}
    \caption{Spectra from the ejecta shock breakout (top left panel), Type Ia SN explosion at peak time (top right panel), magnetar-driven shock breakout (bottom left panel), and magnetar-powered explosion at peak time (bottom right panel). For top panels, the colored dotted, dashed, and solid lines are for fixed $M_{\rm ej} = 1.1,\,1.5,\ {\rm and}\ 2\,M_\odot$ ($M_{\rm Ni}\approx2M_{\rm ej}/3$), respectively. The green and blue lines are for different initial ejecta velocity: $v_{\rm i} = 1.0\ {\rm and}\ 2.0\times10^{4}\,{\rm km}\,{\rm s}^{-1}$, respectively. For bottom panels, the colored dotted, dashed, and solid lines are for fixed $\log_{10}(B/10^{14}\,{\rm G}) = 0.0,\,0.5,\ {\rm and}\ 1.0$, respectively. The blue, green, and orange lines are for different initial period: $P_{\rm i} = 1,\,2\ {\rm and}\ 4\,{\rm ms}$, respectively. The black dashed lines represent the calculated AGN spectrum with the SMBH accreting at near the Eddington luminosity  \citep{hubeny2001}. The SMBH mass is taken as $M_{\rm SMBH} = 10^8\,M_\odot$.  }
    \label{fig:specta}
\end{figure*}

\begin{figure*}[htbp]
    \centering
	\includegraphics[width = 0.49\linewidth , trim = 55 0 60 20 , clip]{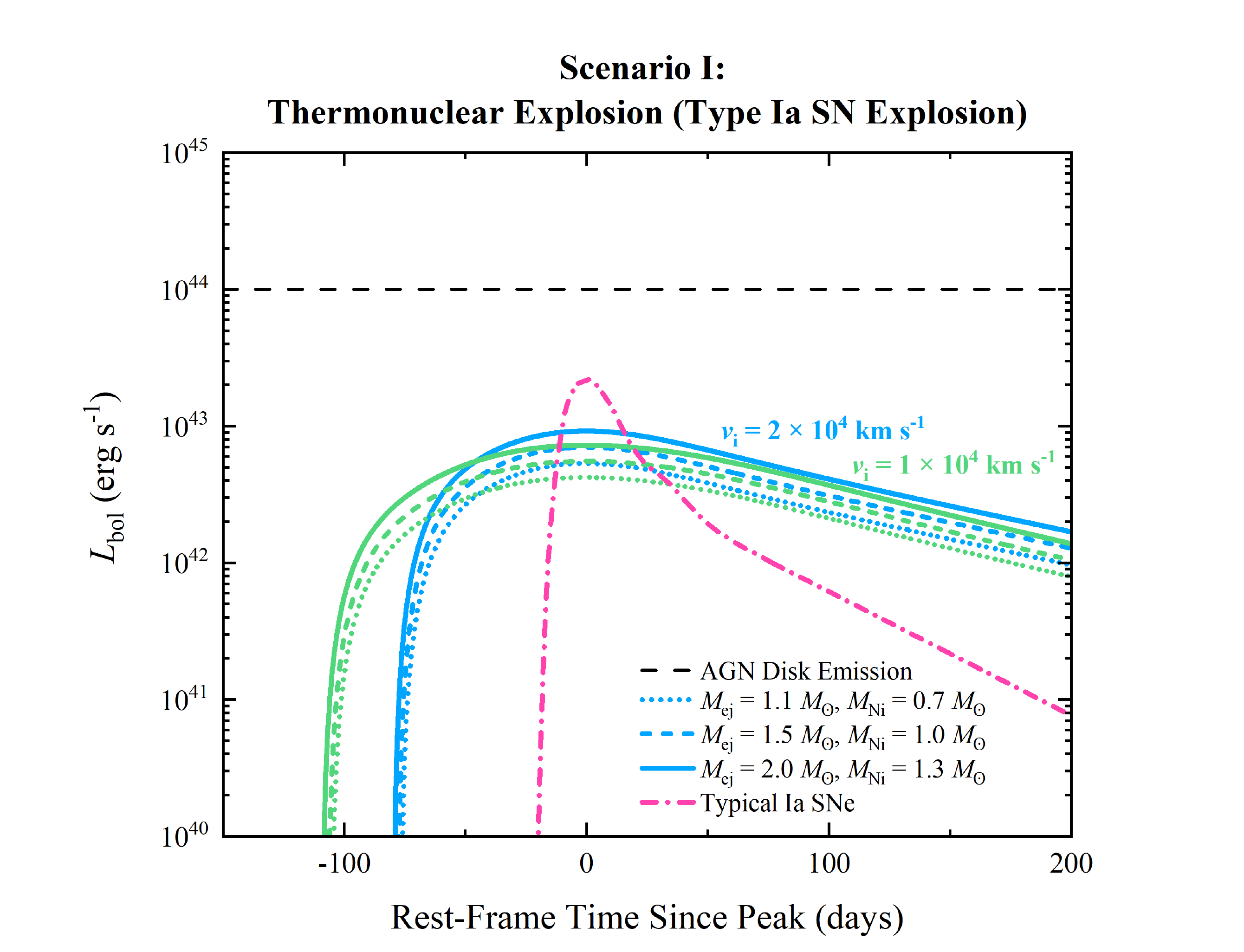}
	\includegraphics[width = 0.49\linewidth , trim = 55 0 60 20 , clip]{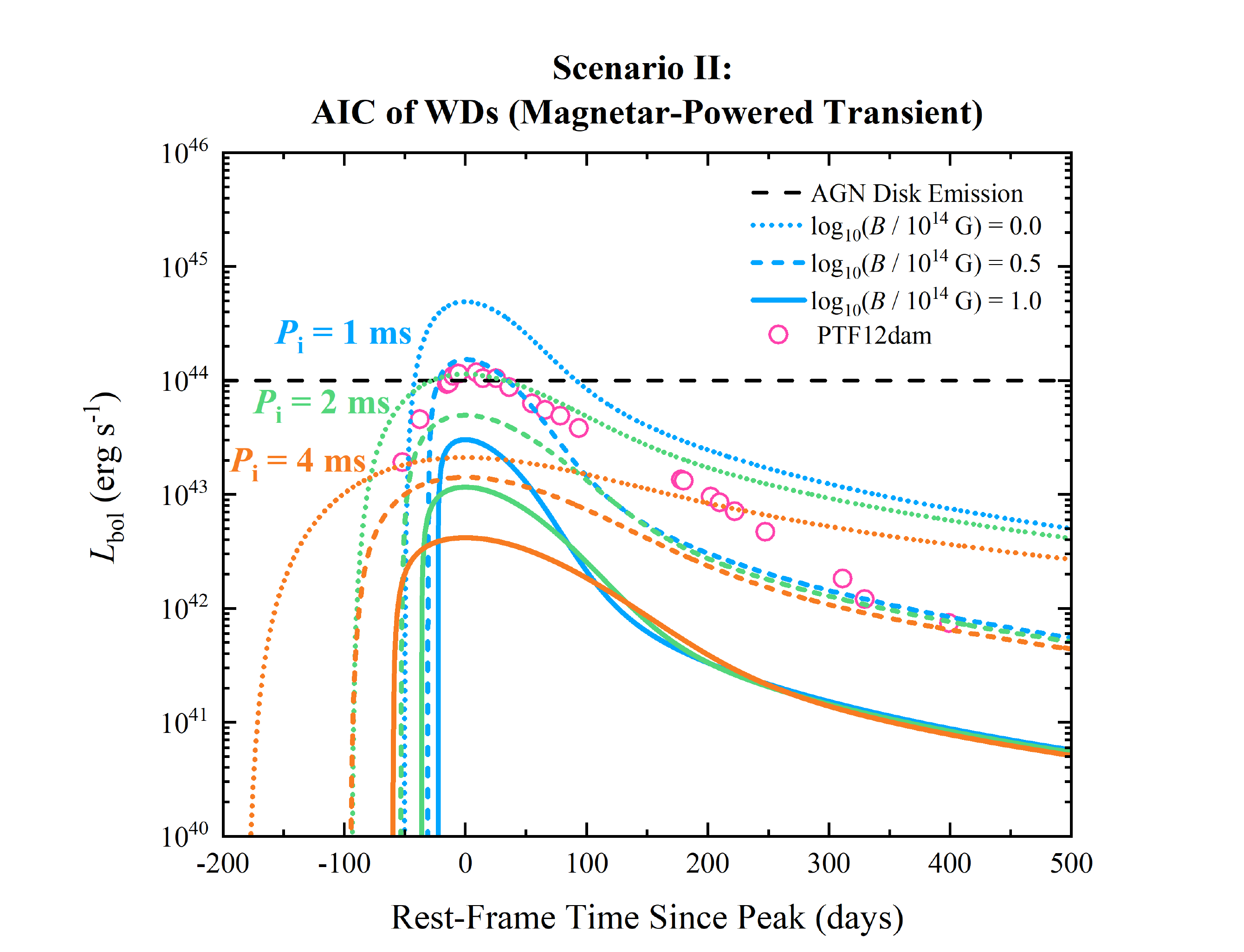}
    \caption{Lightcurves of Type Ia SN explosion (left panel) and magnetar-powered transient (right panel). Labels are similar to those of Figure \ref{fig:lightcurve}. The pink line in the left panel shows the observed lightcurve of a typical SN Ia \citep{branch2017}. The pink circles in the right panel are the observed bolometric data of a typical SLSN-I \citep[PTF12dam,][]{mccrum2014}.   }
    \label{fig:lightcurve}
\end{figure*}

We show the indicative spectra of several types of transients discussed in Section \ref{sec:model} with consideration of different parameters, along with the spectra of a typical AGN accretion disk and its corona \citep{hubeny2001} in Figure \ref{fig:specta}. We also present the lightcurves of a Type Ia SN and a magnetar-powered transient in Figure \ref{fig:lightcurve} by simply adopting the one-dimensional model from \cite{arnett1979} and \cite{kasen2010}. The late-phase $\gamma$-ray escaping effect is ignored. The location of the WD and the mass of SMBH are taken as their typical values as discussed in Section \ref{sec:model}. 

We firstly investigate the effect of ejecta mass $M_{\rm ej}$ and initial velocity $v_{\rm i}$ on the emission from ejecta shock breakout and type Ia SN explosion of scenario I. $M_{\rm ej}$ has little effect on the brightness of shock breakout, since it has little change by a factor of at most two. However, $v_{\rm i}$ can have significant influence on the shock breakout brightness. The ejecta shock breakout signals, brighten in UV to soft X-ray, are always less contaminated by the contribution of AGN emission and hence possible to detect. Type Ia SNe occurring in AGN disks have a much longer rising time compared with classical type Ia SNe. The peak luminosity of a such type Ia SN is always $\sim 8\times 10^{42}\,{\rm erg}\,{\rm s}^{-1}$ in optical, which is an order of magnitude dimmer than the AGN disk emission. Therefore, One can hardly detect the emission from the subsequent Ia SN after the shock breakout signal.

For an AIC event (scenario II), variables include the surface magnetic field $B$ and the initial spin period $P_{\rm i}$ of the NS. Shown in the bottom panels of Figure \ref{fig:specta}, the magnetic-driven shock breakouts tend to be bright while the subsequent magnetar-powered transients tend to be dim when the nascent NSs have faster rotations and strong magnetic fields. This is because high $B$ and low $P_{\rm i}$ would cause the spindown process much more quickly ($t_{\rm sd}\propto B^{-2}P_{\rm i}^2$) before heat diffusion, so that most of the rotational energy of the NS would be transformed into the kinetic energy of the ejecta rather than the radiation energy. If $B \gtrsim 3\times10^{14}\,{\rm G}$, the magnetar usually radiates almost all of its rotational energy before the shock breaks out from the AGN disks. Thus, there is no significant difference between the signals of a magnetar-driven shock breakout if the magnetar rotates with the same initial period and has  $B\gtrsim 3\times10^{14}\,{\rm G}$. 
The calculated AIC events can produce bright magnetar-driven shock breakout signals from UV to soft X-ray bands, which are always brighter than the AGN disk emission. When $B\lesssim 3\times10^{14}\,{\rm G}$, the magnetar-powered explosions would show up as bright SN-like explosions on top of the AGN disk emission in the optical and NUV bands. The peak luminosity, peak time, and lightcurve of these transients resemble those of classical SLSNe. The transients powered by magnetars with $B\gtrsim3\times10^{14}\,{\rm G}$ are outshone by the disk emission and may be difficult to detect.

\subsection{Observational Features in Specific Models of AGN Disks}

\begin{figure*}[htbp]
    \centering
	\includegraphics[width = 0.45\linewidth , trim = 67 45 90 20 , clip]{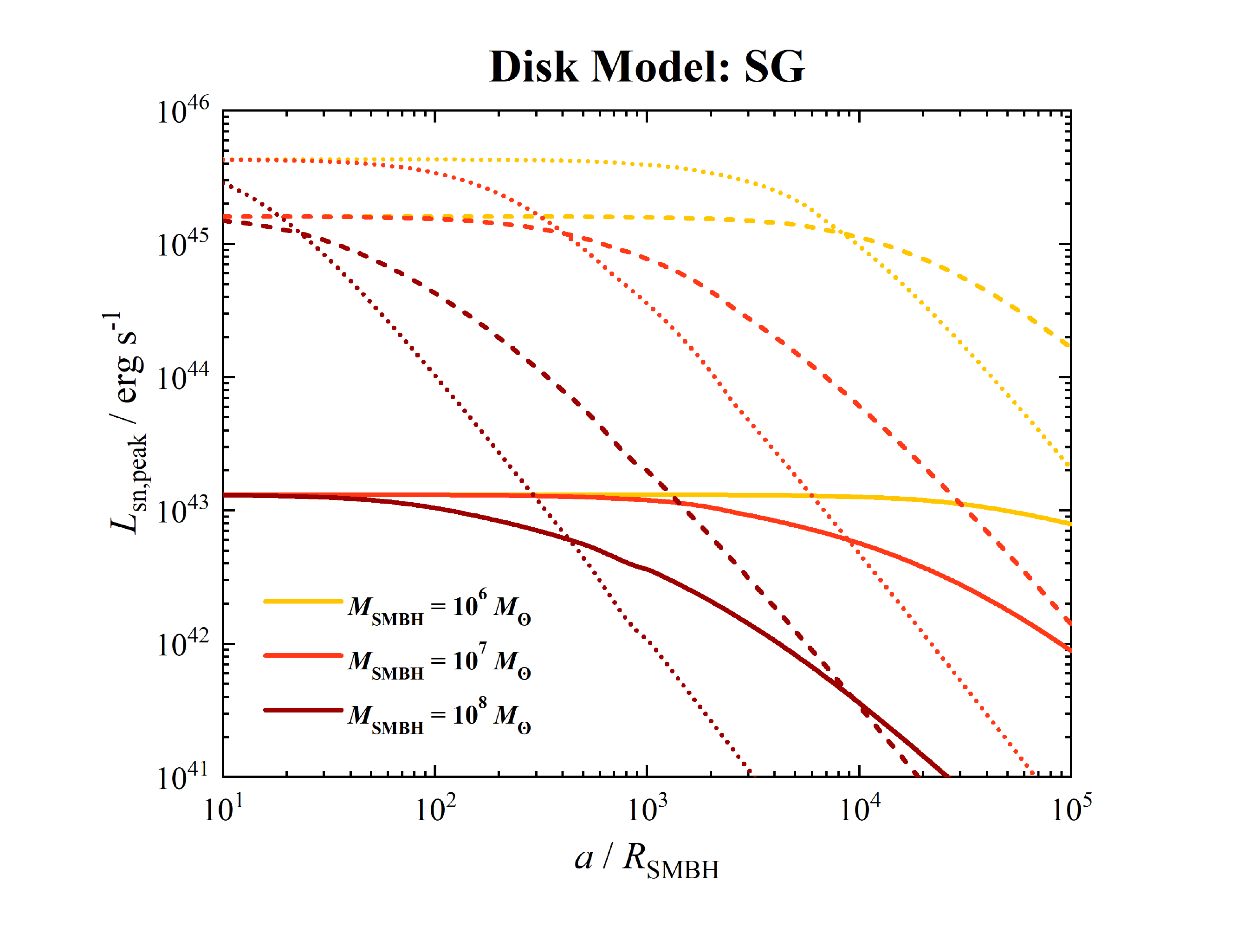}
	\includegraphics[width = 0.45\linewidth , trim = 67 45 90 20 , clip]{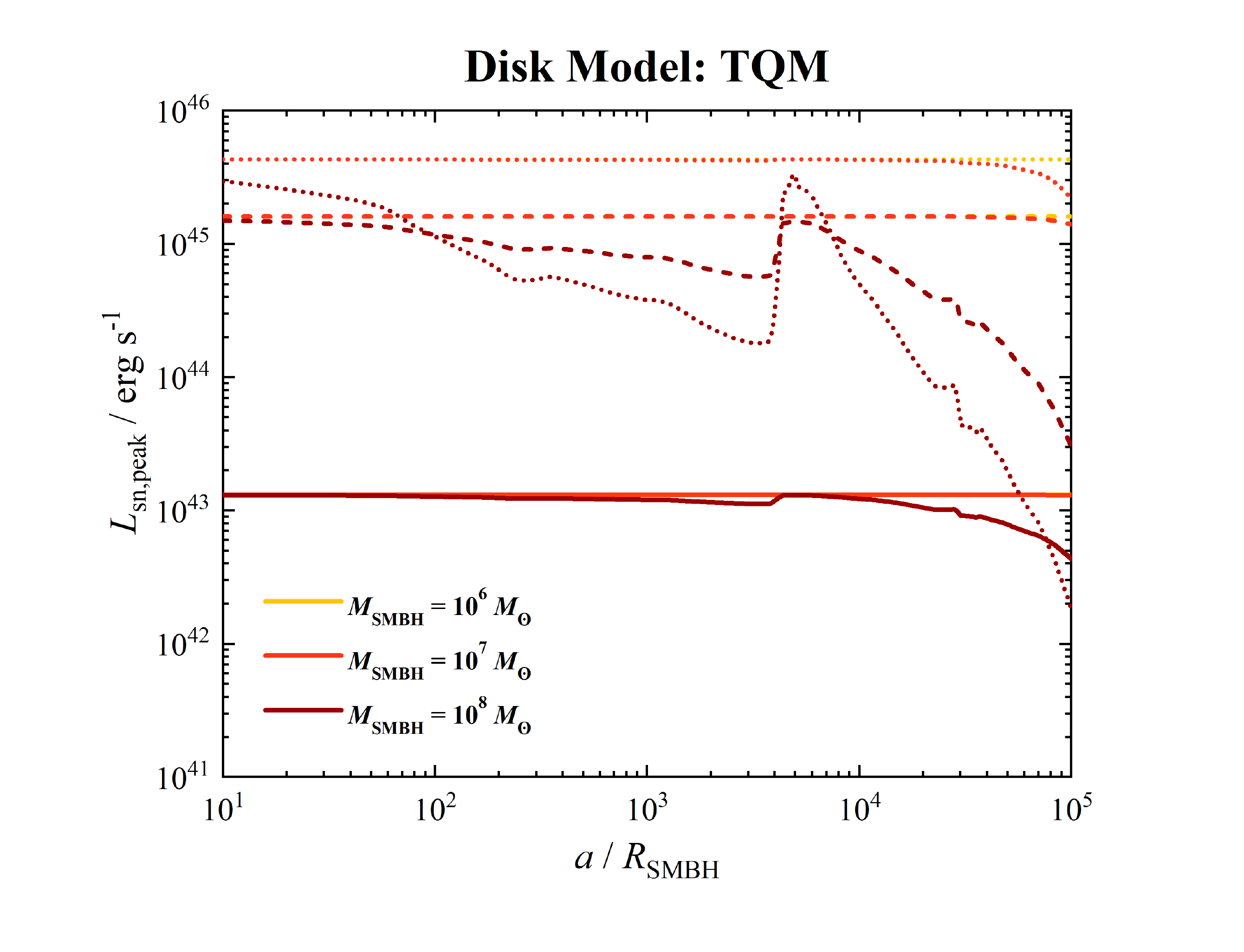}
	\includegraphics[width = 0.45\linewidth , trim = 67 45 90 50 , clip]{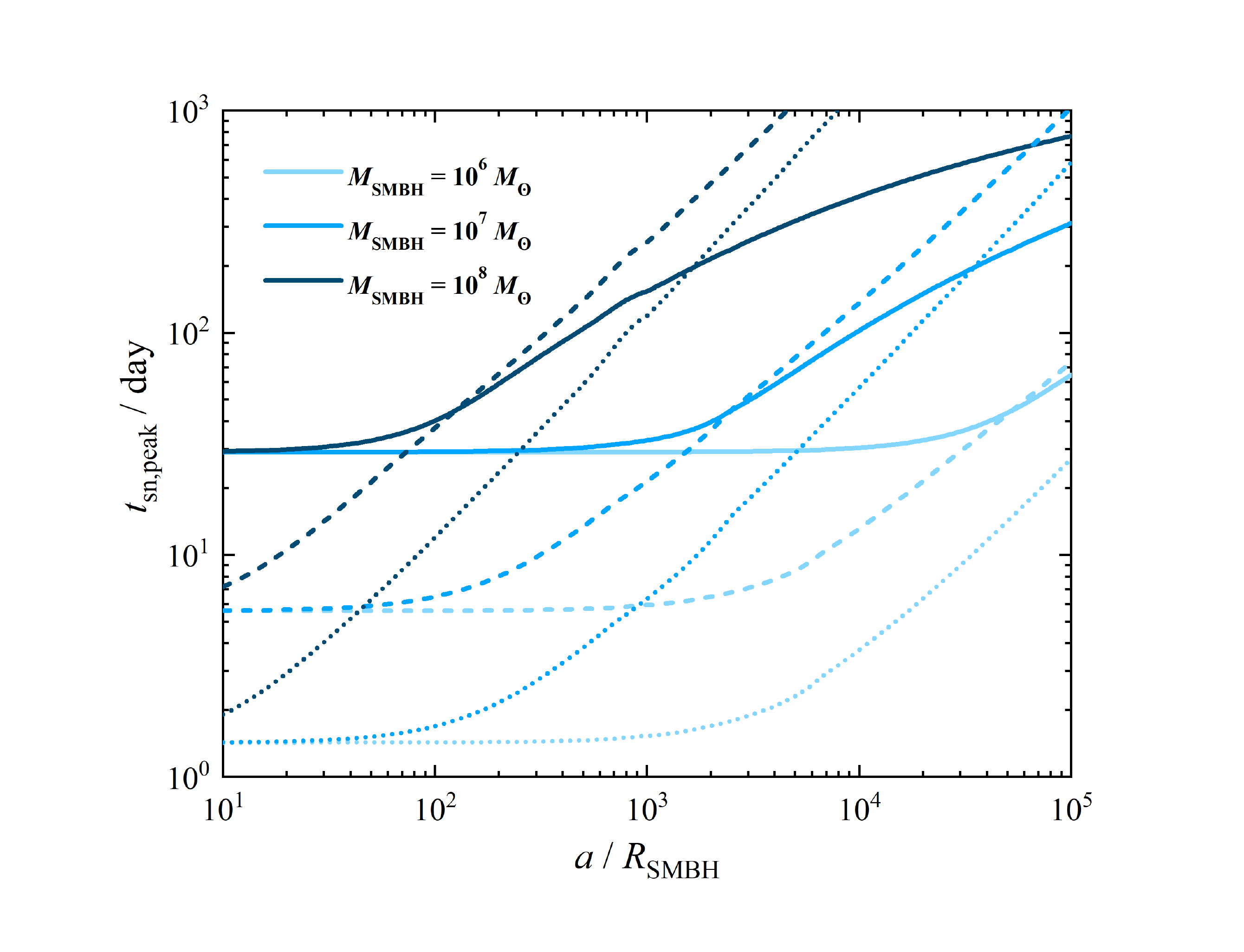}
	\includegraphics[width = 0.45\linewidth , trim = 67 45 90 50 , clip]{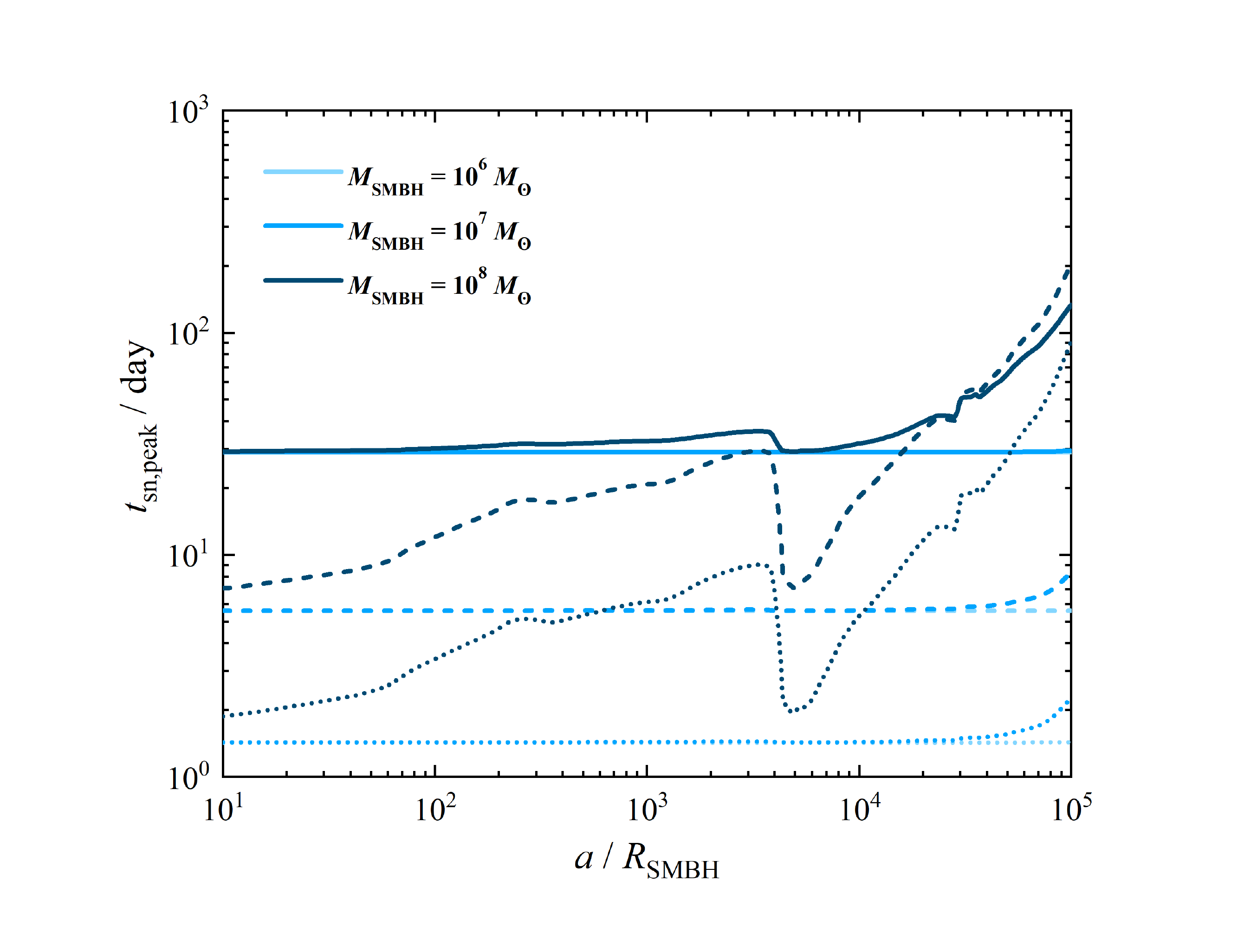}
	\includegraphics[width = 0.45\linewidth , trim = 67 45 90 50 , clip]{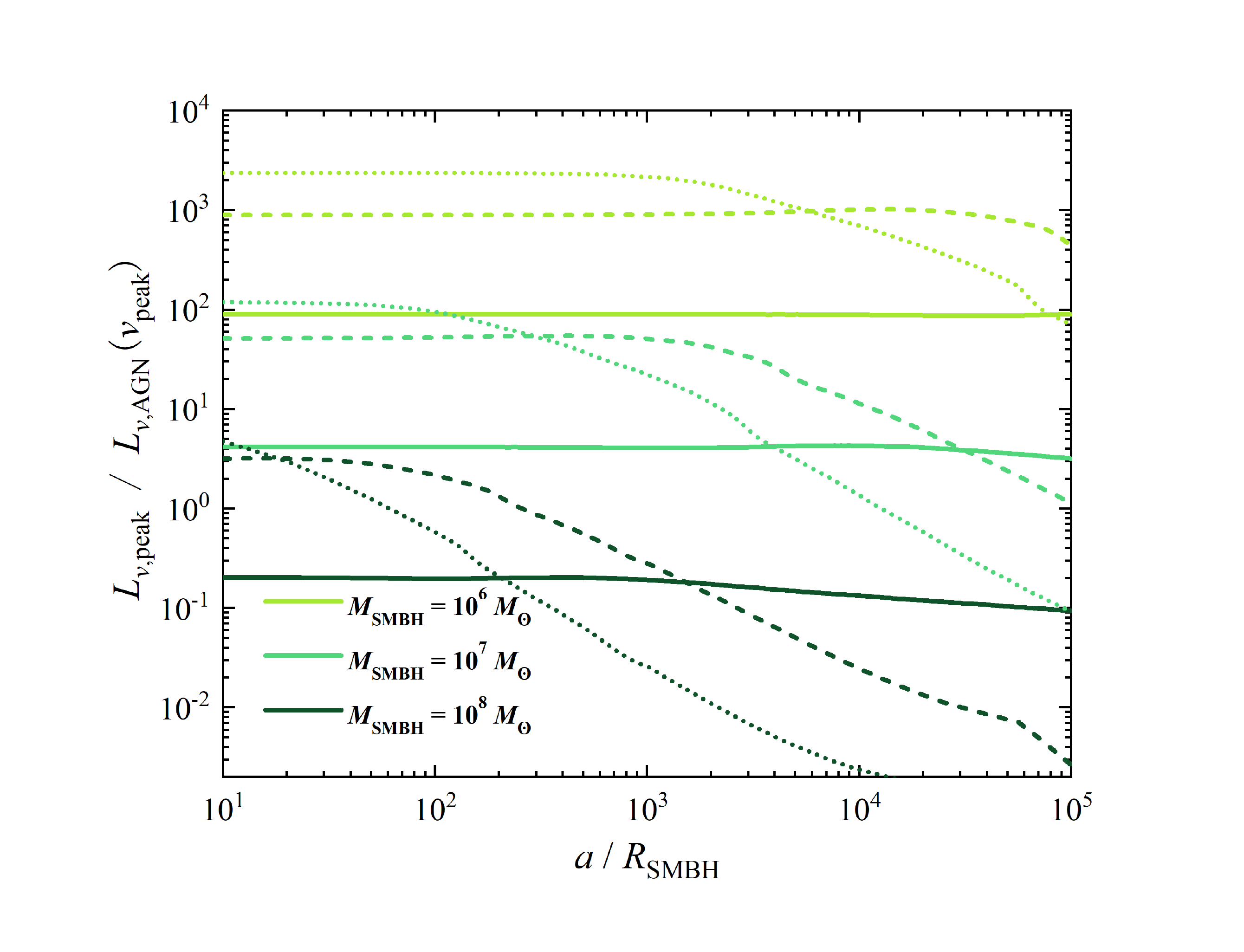}
	\includegraphics[width = 0.45\linewidth , trim = 67 45 90 50 , clip]{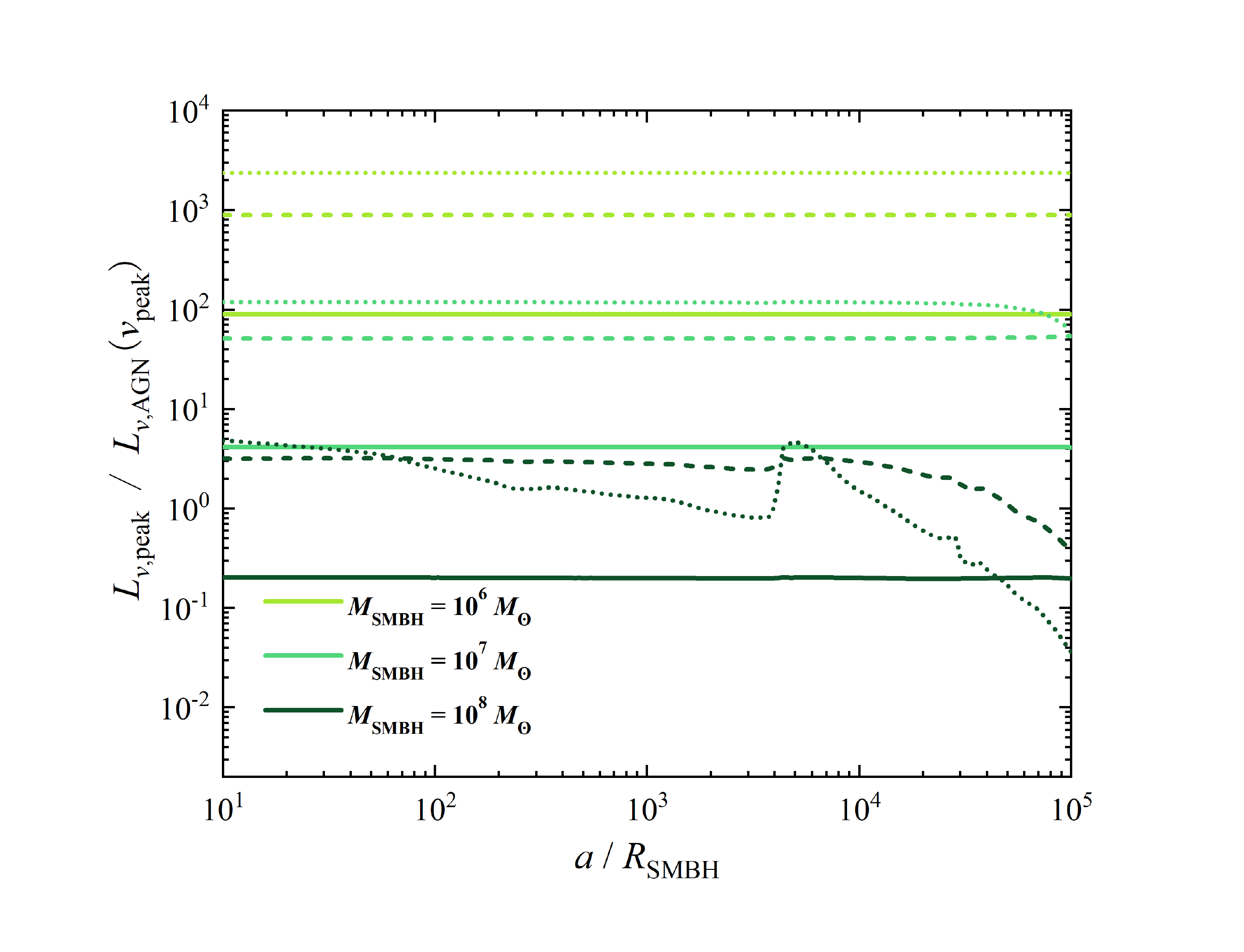}
    \caption{The peak luminosity (top panels), the peak timescale (middle panels), and the brightness ratio (bottom panels) for Type Ia SNe (solid lines) and magnetar-powered explosions (dashed lines and dotted lines) generated in AGN disks, for three values of the SMBH mass (see the labels in each panel). The dashed and dotted lines are for $B = 10^{14}\,{\rm G}$ and $B = 10^{15}\,{\rm G}$, respectively, with the magnetar initial period fixed to $P_{\rm i} = 2\,{\rm ms}$. The density profile and disk scale are adopted from the model of \cite{sirko2003} in the left panels and from the model of \cite{thompson2005} in the right panels. }
    \label{fig:AGNModel}
\end{figure*}

In order to assess different observational features that thermonuclear explosions and AICs of WDs in AGN disks would have, we take into account two specific AGN disk models, including the model by \cite{sirko2003} (hereafter, SG model) and the model by \cite{thompson2005} (hereafter, TQM model). In each model, we consider three values for the mass of SMBH, i.e., $M_{\rm SMBH} = 10^6,~10^7,~{\rm and}~10^8\,M_\odot$. Since shock breakouts are always fast evolving and hard to detect, we only investigate the observational features of Type Ia SNe discussed in scenario I and magnetar-powered explosions discussed in scenario II. The peak luminosity, the peak timescale, and the brightness ratio between the peak luminosity and AGN emission at peak times for these two types of explosions are displayed in Figure \ref{fig:AGNModel}. For AIC of WDs, by setting the mass of AIC ejecta as $M_{\rm ej} = 0.03\,M_\odot$ and the initial spin of the AIC NS as $P_{\rm i} = 2\,{\rm ms}$, two values of the magnetar surface magnetic field strength are adopted, i.e., $B = 10^{14}\,{\rm G}$ and $B = 10^{15}\,{\rm G}$.

We firstly examine the $M_{\rm SMBH} = 10^6\,M_\odot$ case: In both disk models, thermonuclear explosions of WDs within AGN disks can power classical SNe Ia while spindown of AIC NSs can always produce bright and fast-evolving transients. The magnetar-powered transients would be brighter for a stronger magnetic field of the AIC-produced magnetar. Due to the relatively dim brightness of the disk emission, this implies that both explosions are much brighter than the emission from the AGN disks and easily to be detected. 

For the case of a more massive SMBH, $M_{\rm SMBH} = 10^7\,M_\odot$, the observational features of the transients show less changes for the TQM model. Conversely, it takes more time for these transients in the outer parts of the disk ($a / R_{\rm SMBH} \gtrsim 10^3$) to reach the peak by considering the SG disk model. Under this condition, Type Ia SNe would be dimmer than classical Type Ia SNe, and the lightcurves of magnetar-powered transients look like SLSNe. They are always much luminous than the accretion disks of the AGNs.

As the SMBH mass increases to $M_{\rm SMBH} = 10^8\,M_\odot$, Type Ia SNe in AGN disks are always outshone by the AGN disk emission and hence, hardly to be discovered. For the SG model, only AIC events occurring at small radii ($a / R_{\rm SMBH} \lesssim 10^2 $) can be brighter than the AGN disk. However, these magnetar-powered transients could be fast-evolving with a peak time of $\lesssim 10\,{\rm days}$, which makes it difficult to  identify them from various survey projects. For the TQM model, magnetars with $B \sim 10^{14}\,{\rm G}$ would power SLSN-like explosions while magnetars with higher magnetic fields (e.g., $B \sim 10^{15}\,{\rm G}$) can produce fast-evolving transients. They are always bright enough to be discovered. 

\section{Summary and Discussion}

In this paper, we investigate the EM signatures of mass-accreting WDs in AGN disks when they are catastrophically destroyed. Two scenarios are considered: thermonuclear explosion and AIC of WD
into a NS. The thermonuclear explosion could result in an ejecta shock breakout signature from the disk and a Type Ia SN subsequently. The AIC of the WD could lead to the formation of a millisecond magnetar which can deposit its rotational energy into the AGN disk material through spindown. Therefore, a magnetar-driven shock breakout from the disk surface and a bright magnetar-powered transient could emerge from the AGN disk emission. This transient is rapidly evolving and could be luminous if the magnetic field of magnetar is as high as $B\sim10^{15}\,{\rm G}$, while its rising time and peak luminosity could have similar properties with those of SLSNe for $B\sim10^{14}\,{\rm G}$. Type Ia SNe and magnetar-powered transients taking place in the inner parts of the disk around a relatively less massive SMBHs ($M_{\rm SMBH}\lesssim 10^8\,M_\odot$) are more likely to be brighter than the AGN disk emission and hence, more easily to be discovered. {The local mass function of SMBHs show that $\Phi_{M} \sim (0.003 , 0.01)\,{\rm Mpc}^{-3}\,(\Delta \log M)^{-1}$ for SMBHs with $\lesssim10^8\,M_\odot$ while $\Phi_{M}$ decay rapidly for more massive SMBHs  \cite[e.g.,][]{kelly2012}. Therefore, there are a large number of relatively less massive SMBHs as cradles to accommodate AGN mass-accreting WDs in the local Universe. }

Thermonuclear explosions and AICs of WDs in AGN disks are potential sources for future joint gravitational wave (GW), EM, and neutrino multi-messenger observations. Besides AIC of WDs, another possible formation channel of millisecond magnetars in AGN disks is mergers of BNSs. Such a merger may produce a millisecond magnetar if the NS equation of state is stiff enough \citep[e.g.][]{ai20}. This would also drive a magnetar-powered explosion as discussed above. The mechanism by which a nascent millisecond magnetar might power a GRB jet has been studied in detail over the past decade \citep{bucciantini2008,bucciantini2012,metzger2008,metzger2011,siegel2014}\footnote{Some indirect evidence for the magnetar mechanism in GRBs are the presence of a plateau in the early X-ray afterglow of long-duration GRBs, which may be explained as the continuous energy injection from the spindown of a magnetar \citep{dai1998,zhang2001,zhang2006}.}. A putative jet may thus be launched from the AIC of WDs and BNS mergers occurring in AGN disks. However, \cite{zhu2021a,zhu2021b,perna2021a} recently showed that GRB jets embedded in AGN disks would be usually choked by the dense material of the accretion disks. Although it is hard to observe  gamma-ray signals, \cite{zhu2021b} suggested that these choked GRBs may effectively produce TeV–PeV neutrinos that could be detected by IceCube and IceCube-Gen2. Furthermore, the dissipation of magnetic energy during the merger of BWDs is expected to accelerate cosmic rays and produce high energy neutrinos \citep{xiao2016}. In the GW channel, BWD mergers are promising astrophysical GW sources for space-borne GW observatories, e.g., LISA \citep{amaroseoane2017}, TaiJi \citep{ruan2018}, and TianQin \citep{luo2016}, while the GW signals from BNS mergers are readily detected with the ground-based Gw detectors such as LIGO and Virgo \citep{abbott2017}. Future joint observations of GW, EM, and neutrino signals can reveal the existence of WD and NS populations in AGN disks.

The event rate of thermonuclear explosions and AICs by WDs in the AGN channel is amphibious at present. {Since the lifetime of less-massive stars is much longer than the lifetime of the AGNs, AGN WDs may be captured  from the nuclear star clusters around the SMBHs.} \cite{mckernan2020} simulated the relative rate of BWD mergers in AGN disks and obtained $f_{\rm AGN}[0.2 , 5000]\,{\rm Gpc}^{-3}\,{\rm yr}^{-1}$, {where $f_{\rm AGN}\sim10\%-50\%$ \citep{yang2019b} is defined as the fraction of the LIGO-Virgo observed binary BH mergers  originating from AGNs. This gives the range of binary WD mergers in AGN disks as $\sim [0.02, 2500]\,{\rm Gpc}^{-3}\,{\rm yr}^{-1}$. } \cite{perna2021b} computed the rates of AICs of NSs to BHs for both isolated NSs and NSs formed from BNS mergers, which are $\sim [0.07 , 20]\,{\rm Gpc}^{-3}\,{\rm yr}^{-1}$, respectively. Considering that the number of WDs in AGN disks could be comparable to the number of AGN NSs \citep{mckernan2020}, one may estimate that the rate of thermonuclear explosions and AICs of AGN WDs could be approximately up to this rate. We would like to further study specific event rates for thermonuclear explosions and AICs of WDs in the future.
 
Type Ia SNe and magnetar-powered explosions within AGN disks could have some unique observational features. After the ejecta passes through the disk photosphere, the disk materials in principle could still be dense although the disk has an exponentially vertical decaying density profile. Ejecta-disk matter interaction could possibly take place at the early stage of the thermonuclear explosions and AICs of WDs in AGN disks. Such an effect could provide an additional energy source that makes the explosions brighter and cause the early spectra of the transients to look like SNe IIn \citep[e.g.,][]{chatzopoulos2014,smith2014,liu2018}. If ejecta could expand into the broad-line regions, the interaction between ejecta and clouds in broad-line regions can drive SN IIn-like transients \citep[e.g.,][]{moriya2017}\footnote{The width of hydrogen lines may be some what wider than Type IIn SNe in view that the interaction occurs in the broad line region.}. The dust torus could possibly cover the emission from explosions along the line of sight so that the observed emission in infrared can be dominated by dust \citep[e.g.,][]{assef2018}. Recently, \cite{jermyn2021} suggested that AGN stars can have near-critical spins and undergo quasi-chemically-homogeneous evolution, which share the similarity with the Wolf-Rayet progenitors of standard long-duration GRBs. Besides, magnetar-powered SLSNe are expected to be the products of rapidly rotating Wolf–Rayet stars similar to LGRBs \citep[e.g.,][]{lunnan2014,yu2017,aguileradena2018,blanchard2020}, so SLSNe could potentially occur in AGN disks.  SLSN candidates within the central regions of AGNs were reported by \cite{drake2011,assef2018}. Magnetar-powered explosions via AIC of AGN WDs discussed in this paper and magnetar-powered SLSNe can have similar peak brightnesses, peak times, and lightcurve evolution patterns. However, before the magnetic-driven shock breakout, the collapse of a star in an AGN disk would eject its envelope and cause an initial ejecta shock breakout from the AGN disk surface. One can observe two shock breakouts, i.e., the initial ejecta shock breakout and a second magnetic-driven shock breakout, before the main peak of an AGN-star-powered SLSN. Furthermore, for the magnetar-powered explosions via AIC of AGN WDs, the absorption features of carbon and oxygen in the spectra of classical SLSN-I \citep[see][for a review]{galyam2019} could disappear due to the lack of carbon and oxygen envelopes. These different observational features may be used to distinguish magnetar-powered explosions due to AIC events and classical SLSNe embedded in the AGN disks. 

\acknowledgments
We thank Ping Chen and Ming-Yang Zhuang for valuable comments, and an anonymous referee for constructive suggestions. The work of J.P.Z is partially supported by the National Science Foundation of China under Grant No. 11721303 and the National Basic Research Program of China under grant No. 2014CB845800. Y.P.Y is supported by National Natural Science Foundation of China grant No. 12003028 and Yunnan University grant No.C176220100087. L.D.L. is supported by the National Postdoctoral Program for Innovative Talents (grant No. BX20190044), China Postdoctoral Science Foundation (grant No. 2019M660515), and “LiYun” postdoctoral fellow of Beijing Normal University. Y.W.Y is supported by the National Natural Science Foundation of China under Grant No. 11822302, 11833003. H.G. is supported by the National Natural Science Foundation of China under Grant No. 11722324, 11690024, 11633001, the Strategic Priority Research Program of the Chinese Academy of Sciences, Grant No. XDB23040100 and the Fundamental Research Funds for the Central Universities. 

\bibliography{WD_in_AGN}{}
\bibliographystyle{aasjournal}

\end{document}